\documentclass{article}
\usepackage{amssymb}
\usepackage{amsmath}
\usepackage{theorem}
\usepackage{graphicx}

\newtheorem{property}{Property}
\newtheorem{lemma}{Lemma}
\newtheorem{corollary}{Corollary}
\newenvironment{proof}[1][Proof]{\noindent\textbf{#1.} }{\ \rule{0.5em}{0.5em}}
\newenvironment{definition}[1][Definition.]{\begin{trivlist}
\item[\hskip \labelsep {\bfseries #1}]}{\end{trivlist}}
\newenvironment{theorem}[1][Theorem]{\begin{trivlist}
\item[\hskip \labelsep {\bfseries #1}]}{\end{trivlist}}
\newenvironment{acknowledgement}[1][Acknowledgement.]{\begin{trivlist}
\item[\hskip \labelsep {\bfseries #1}]}{\end{trivlist}}

\title{On Edge-Disjoint Pairs Of Matchings\thanks{%
The work on this paper was supported by a grant of the Armenian
National Science and Educational Fund.}}
\author{V. V. Mkrtchyan$^{\dagger}$, V. L. Musoyan$^{\ddagger}$, A. V.
Tserunyan$^{\S }$}

\date{}

\begin{document}
\maketitle

\begin{center}
Department of Informatics and Applied Mathematics, Yerevan State
University, Yerevan, 0025, Republic of Armenia

$^{\dagger}$Institute for Informatics and Automation Problems of
National Academy of Sciences of Armenia

$^{\S }$ Department of Mathematics, University of California at Los
Angeles, CA 90095, USA

$^{\dagger}$vahanmkrtchyan2002@\{ysu.am, ipia.sci.am, yahoo.com\}

$^{\ddagger}$ vahe\_musoyan@ysu.am, vahe.musoyan@gmail.com

$^{\S }$ tserunyan@ucla.edu, anush\_tserunyan@yahoo.com

\bigskip

\textit{Dedicated to the 35th anniversary of Discrete Mathematics}
\end{center}

\begin{abstract}
For a graph $G$, consider the pairs of edge-disjoint matchings whose union
consists of as many edges as possible. Let $H$ be the largest matching among
such pairs. Let $M$ be a maximum matching of $G$. We show that $5/4$ is a
tight upper bound for $|M|/|H|$.\newline
\end{abstract}


We consider finite, undirected graphs without multiple edges or loops. Let $%
V(G)$ and $E(G)$ denote the sets of vertices and edges of a graph $G$,
respectively. The cardinality of a maximum matching of a graph $G$ is
denoted by $\nu (G)$.

For a graph $G$ define $B_{2}(G)$ as follows:

\begin{center}
$B_{2}(G)\equiv \{(H,H^{\prime }):H,H^{\prime }$ are edge-disjoint matchings
of $G\}$,
\end{center}

and set:

\begin{center}
$\lambda_2 (G)\equiv \max \{\left\vert H\right\vert +\left\vert H^{\prime
}\right\vert :(H,H^{\prime })\in B_{2}(G)\}$.
\end{center}

Define:

\begin{center}
$\alpha_2 (G)\equiv \max \{\left\vert H\right\vert ,\left\vert H^{\prime
}\right\vert :$ $(H,H^{\prime })\in B_{2}(G)$ and $\left\vert H\right\vert
+\left\vert H^{\prime }\right\vert =\lambda_2 (G)\}$,

$M_{2}(G)\equiv \{(H,H^{\prime }):(H,H^{\prime })\in B_{2}(G),\left\vert
H\right\vert +\left\vert H^{\prime }\right\vert =\lambda_2 (G),\left\vert
H\right\vert =\alpha_2 (G)\}$.
\end{center}

It is clear that $\alpha_2(G) \leq \nu(G)$ for all $G$. By
Mkrtchyan's result [4], reformulated as in [2], if $G$ is a matching
covered tree then the inequality turns to an equality. Note that a
graph is said to be matching covered (see [5]) if its every edge
belongs to a maximum matching (not necessarily a perfect matching as
it is usually defined, see e.g. [3]).

The aim of this paper is to obtain a tight upper bound for $\frac{\nu (G)}{%
\alpha_2 (G)}$. We prove that $\frac{5}{4}$ is an upper bound for $\frac{\nu
(G)}{\alpha_2 (G)}$, and exhibit a family of graphs which shows that $\frac{5%
}{4}$ can not be replaced by any smaller constant. Terms and
concepts that we do not define can be found in [1, 3, 6].

\bigskip

Let $A$ and $B$ be matchings of a graph $G$.

\begin{definition}
\label{altpaths} \textit{A path (or an even cycle) $e_1, e_2, ...,
e_l\ (l\geq1)$ is called $A$-$B$ alternating if the edges with odd
indices belong to $A \backslash B$ and others to $B \backslash A$,
or vice-versa.}
\end{definition}

\begin{definition}
\label{maxaltpaths} \textit{$A$-$B$ alternating path $P$ is called maximal
if there is no other $A$-$B$ alternating path that contains $P$ as a proper
subpath.}
\end{definition}

The sets of $A$-$B$ alternating cycles and maximal alternating paths are
denoted by $C(A,B)$ and $P(A,B)$, respectively.

The set of the paths from $P(A,B)$ that have even (odd) length is denoted by
$P_e(A,B)$ ($P_o(A,B)$).

The set of the paths from $P_o(A,B)$ starting from an edge of $A$ ($B$) is
denoted by $P_o^A(A,B)$ ($P_o^B(A,B)$).

Note that every edge $e\in A\cup B$ either belongs to $A\cap B$ or lies on a
cycle from $C(A,B)$ or lies on a path from $P(A,B)$.

Moreover, \renewcommand{\labelenumi}{(\alph{enumi})}

\begin{property}\label{AB}
\begin{enumerate}
\item if $F \in C(A,B) \cup P_e(A,B)$ then $A$ and $B$ have the same number of edges that
belong to $F$,

\item if $P\in P_o^A(A,B)$ then the difference between
the numbers of edges that lie on $P$ and belong to $A$ and $B$ is
one.
\end{enumerate}
\end{property}

These observations immediately imply:

\begin{property}
\label{cardinalitydiff} If $A$ and $B$ are matchings of a graph $G$ then
\begin{equation*}
|A|-|B|=|P_{o}^{A}(A,B)|-|P_{o}^{B}(A,B)|.
\end{equation*}
\end{property}

Berge's well-known theorem states that a matching $M$ of a graph $G$
is maximum if and only if $G$ does not contain an $M$-augmenting
path [1,3,6]. This theorem immediately implies:

\begin{property}
\label{maxmatchingproperty} If $M$ is a maximum matching and $H$ is a
matching of a graph $G$ then
\begin{equation*}
P_{o}^{H}(M,H)=\emptyset,
\end{equation*}%
and therefore, $|M|-|H|=|P_{o}^{M}(M,H)|$.
\end{property}

The proof of the following property is similar to the one of property \ref%
{maxmatchingproperty}:

\begin{property} \label{HH'}
If $(H,H')\in M_2(G)$ then $P_o^{H'}(H,H')=\emptyset$.
\end{property}

Let $G$ be a graph. Over all $(H,H^{\prime })\in M_{2}(G)$ and all maximum
matchings $M$ of $G$, consider the pairs $((H,H^{\prime }),M)$ for which $%
|M \cap H|$ is maximized. Among these, choose a pair $%
((H,H^{\prime }),M)$ such that $|M \cap H'|$ is maximized.

From now on $H,H^{\prime}$ and $M$ are assumed to be chosen as described
above. For this choice of $H,H^{\prime}$ and $M$, consider the paths from $%
P_o^M(M,H)$ and define $M_A$ and $H_A$ as the sets of edges lying on these
paths that belong to $M$ and $H$, respectively.

\begin{lemma}
\label{MHaltpaths} $C(M,H)=P_{e}(M,H)=P_{o}^{H}(M,H)=\emptyset $.
\end{lemma}

\begin{proof}
Property \ref{maxmatchingproperty} implies $P_{o}^{H}(M,H)=\emptyset$.
Let us show that $C(M,H)=P_{e}(M,H)=\emptyset $. Suppose that there is $%
F_{0}\in C(M,H)\cup P_{e}(M,H)$. Define:
\begin{equation*}
M^{\prime }\equiv \lbrack M\backslash E(F_{0})]\cup \lbrack H\cap E(F_{0})].
\end{equation*}

Consider the pair $((H,H^{\prime }),M^{\prime })$. Note that $M'$ is
a maximum matching, and
\begin{equation*}
|H \cap M'| > |H \cap M|,
\end{equation*}%

which contradicts $|H \cap M|$ being maximum.
\end{proof}

\begin{corollary}\label{outerHMs}
$M \cap H = M \backslash M_A = H \backslash H_A$.
\end{corollary}

\begin{lemma}
\label{M_AandH's} Each edge of $\ M_{A}\backslash H^{\prime }$ is adjacent
to two edges of $H^{\prime }$.
\end{lemma}

\begin{proof}
Let $e$ be an arbitrary edge from $M_{A}\backslash H^{\prime }$. Note that $%
e\in M$, $e\notin H$, $e\notin H^{\prime }$. Now, if $e$ is not adjacent to
an edge of $H^{\prime }$, then $H\cap (H^{\prime }\cup \{e\})=\emptyset $
and
\begin{equation*}
|H|+|H^{\prime }\cup \{e\}|>|H|+|H^{\prime }|=\lambda _{2}(G),
\end{equation*}%
which contradicts $(H,H') \in M_2(G)$.

On the other hand, if $e$ is adjacent to only one edge $f\in
H^{\prime }$, then consider the pair $(H,H^{\prime \prime })$, where
$H^{\prime \prime }\equiv (H^{\prime }\backslash \{f\}) \cup \{e\}$.
Note that
\begin{equation*}
H \cap H^{\prime \prime }=\emptyset, \ |H''| = |H'|
\end{equation*}
and
\begin{equation*}
|H'' \cap M| > |H' \cap M|,
\end{equation*}
which contradicts $|H' \cap M|$ being maximum.
\end{proof}

\begin{lemma}
\label{onlyodd} $C(M_A, H^{\prime}) = P_e(M_A, H^{\prime}) = P_o^{M_A}(M_A,
H^{\prime}) = \emptyset$.
\end{lemma}

\begin{proof}
First of all, let us show that $C(M_A,H^{\prime}) = P_e(M_A, H^{\prime}) =
\emptyset$. For the sake of contradiction, suppose that there is $F_{0}\in
C(M_A, H^{\prime}) \cup P_e(M_A, H^{\prime})$. Define:
\begin{equation*}
H^{\prime\prime}\equiv [ H^{\prime}\backslash E(F_0)] \cup [ M_A \cap
E(F_0)].
\end{equation*}

Consider the pair of matchings $(H,H^{\prime \prime })$. Note that
due to the definition of an alternating path we have $M_{A}\cap
H=\emptyset $, therefore
\begin{equation*}
H\cap H^{\prime \prime }=\emptyset,
\end{equation*}
\begin{equation*}
|H| + |H^{\prime\prime}| = |H| + |H^{\prime}| = \lambda _{2}(G)
\end{equation*}
(see (a) of property \ref{AB}).

Thus $(H,H^{\prime \prime })\in M_{2}(G)$ and%
\begin{equation*}
|H'' \cap M| > |H' \cap M|,
\end{equation*}%
which contradicts $|H' \cap M|$ being maximum.

On the other hand, the end-edges of a path from $P_o^{M_A}(M_A,
H^{\prime})$ are from $M_A$ and are adjacent to only one edge from
$H^{\prime}$ contradicting lemma \ref{M_AandH's}. Therefore,
$P_o^{M_A}(M_A, H')=\emptyset$.
\end{proof}

\begin{lemma}
\label{equality} $|H^{\prime }|=|P_{o}^{H^{\prime }}(M_{A},H^{\prime
})|+|H_{A}|+\nu (G)-\alpha _{2}(G)$.
\end{lemma}

\begin{proof}
Due to property \ref{cardinalitydiff}
\begin{equation*}
|H^{\prime }|-|M_{A}|=|P_{o}^{H^{\prime }}(M_{A},H^{\prime
})|-|P_{o}^{M_{A}}(M_{A},H^{\prime })|,
\end{equation*}%
and due to (b) of property \ref{AB} and property
\ref{maxmatchingproperty}
\begin{equation*}
|M_{A}|-|H_{A}|=|P_{o}^{M}(M,H)|=|M|-|H|=\nu (G)-\alpha _{2}(G).
\end{equation*}%
By lemma \ref{onlyodd} $P_{o}^{M_{A}}(M_{A},H^{\prime })=\emptyset
$, therefore,
\begin{equation*}
|H^{\prime }|=|P_{o}^{H^{\prime }}(M_{A},H^{\prime
})|+|M_{A}|=|P_{o}^{H^{\prime }}(M_{A},H^{\prime })|+|H_{A}|+\nu (G)-\alpha
_{2}(G).
\end{equation*}
\end{proof}

\begin{lemma}
\label{MHpaths} Let $P\in P_{o}(M,H)$ and assume that $%
P=m_1, h_1, m_2, ..., h_{l-1}, m_l$, $l \geq 1, m_i \in M, 1 \leq i
\leq l, h_j \in H, 1 \leq j \leq l-1$. Then $l \geq 3$ and $\{ m_1,
m_l \} \subseteq H'$.
\end{lemma}
\begin{proof}
Let us show that $m_1, m_l \in H'$. If $l = 1$ then $P = m_1$, $m_1
\in M \backslash H$, and $m_1$ is not adjacent to an edge from $H$
as $P$ is maximal. Thus, $m_1 \in H'$ as otherwise we could enlarge
$H$ by adding $m_1$ to it which contradicts $(H, H') \in M_2(G)$.
Thus suppose that $l \geq 2$. Let us show that $m_1 \in H'$. If $m_1
\notin H'$ then define
$$H_1 \equiv (H \backslash \{h_1\}) \cup \{m_1\}.$$

Clearly, $H_1$ is a matching, $H_1 \cap H' = \emptyset$ and $|H_1| =
|H|$, which means that $(H_1, H') \in M_2(G)$. But
$$|H_1 \cap M| > |H \cap M|,$$
which contradicts $|H \cap M|$ being maximum. Similarly, it could be
shown that $m_l \in H'$.

Now let us show that $l\geq 3$. Due to property \ref{HH'},
$P_o^{H'}(H,H') = \emptyset $. Thus there is $i,$ $1\leq i\leq l$,
such that $m_i \in M \backslash H'$. Since $\{m_1, m_l\} \subseteq
H'$, we have $l \geq 3$.
\end{proof}

\begin{corollary}
\label{HAbound} $\left\vert H_{A}\right\vert \geq 2(\nu (G)-\alpha _{2}(G))$.
\end{corollary}

\begin{proof}
Due to lemma \ref{MHpaths} every path $P\in P_{o}(M,H)$ has length at least
five, therefore it contains at least two edges from $H$. Due to property \ref%
{maxmatchingproperty}, there are
$$|P_{o}(M,H)| = |P_{o}^{M}(M,H)| = \nu (G)-\alpha _{2}(G)$$
paths from $P_{o}(M,H)$, therefore
\begin{equation*}
\left\vert H_{A}\right\vert \geq 2(\nu (G)-\alpha _{2}(G)).
\end{equation*}
\end{proof}

\begin{corollary}
\label{VerticesMH} Every vertex lying on a path from $P(M,H)=P_{o}^{M}(M,H)$
is incident to an edge from $H^{\prime }$.
\end{corollary}

\begin{proof}
Suppose $w$ is a vertex lying on a path from $P(M,H)=P_{o}^{M}(M,H)$
and assume that $e$ is an edge from $M_{A}$ incident to the vertex
$w$. Clearly,
if $e\in H^{\prime }$ then the corollary is proved therefore we may assume that $%
e\notin H^{\prime }$. Note that $e\in M_{A}\backslash H^{\prime }$ therefore
due to lemma \ref{M_AandH's} $e$ is adjacent to two edges from $H^{\prime }$%
. Thus $w$ is incident to an edge from $H^{\prime }$.
\end{proof}

\vspace*{0.5cm}

Let $Y$ denote the set of the paths from $P(H, H')$ starting from
the end-edges of the paths from $P_o^M(M, H)$. Note that $Y$ is
well-defined since due to lemma \ref{MHpaths} these end-edges belong
to $H'$. According to property \ref{HH'}, $Y \subseteq P_e(H, H')$,
thus the set of the last edges of the paths from $Y$ is a subset of
$H$. Let us denote it by $H_Y$.

\begin{lemma}\label{H'HpathsandPoddMAH'}
\
\renewcommand{\labelenumi}{(\alph{enumi})}

\begin{enumerate}
\item\label{H'Hpaths} $|Y| = 2(\nu(G) - \alpha_2(G))$ and the length of the paths
from $Y$ is at least four,

\item\label{PoddMAH'} $\left\vert P_{o}^{H^{\prime }}(M_{A},H^{\prime
})\right\vert \geq \nu (G)-\alpha _{2}(G).$
\end{enumerate}
\end{lemma}

\begin{proof}
\textit{(a)} Due to property \ref{HH'}, all end-edges of the paths
from $P_o^M(M,H)$ lie on different paths from $Y$. Therefore $|Y| =
2|P_o^M(M, H)| = 2(\nu(G) - \alpha_2(G))$.

Since the edges from $H_Y$ are adjacent to only one edge from $H'$,
we conclude that they do not lie on a path from $P_o^M(M,H)$
(corollary \ref{VerticesMH}). Thus, due to corollary \ref{outerHMs},
$H_Y \subseteq M\cap H$. Furthermore, as the first two edges of a
path from $Y$ lie on a path from $P_o^M(M,H)$, and the last edge
does not, we conclude that its length is at least four.

\textit{(b)} From $H_Y \subseteq M\cap H$ we get
\begin{equation*}
\left\vert M\cap H\right\vert \geq |H_Y| = |Y| = 2|P_o^M(M, H)| =
2(\nu (G)-\alpha _{2}(G)).
\end{equation*}

On the other hand, every edge from $H_Y$ is adjacent to an edge from $%
H^{\prime }\backslash M$, which is an end-edge of a path from $%
P_o^{H'}(M_A, H')$, therefore%
\begin{equation*}
2(\nu (G)-\alpha _{2}(G))\leq \left\vert M\cap H\right\vert \leq 2\left\vert
P_{o}^{H^{\prime }}(M_{A},H^{\prime })\right\vert
\end{equation*}

or%
\begin{equation*}
\nu (G)-\alpha _{2}(G)\leq \left\vert P_{o}^{H^{\prime }}(M_{A},H^{\prime
})\right\vert .
\end{equation*}
\end{proof}

\bigskip \bigskip

\begin{theorem}
\textit{For every graph $G$ the inequality $\frac{\nu (G)}{\alpha _{2}(G)}\leq \frac{%
5}{4}$ holds.}
\end{theorem}

\begin{proof}
Lemma \ref{equality}, statement \textit{(b)}
of lemma \ref{H'HpathsandPoddMAH'} and corollary \ref{HAbound} imply%
\begin{equation*}
\alpha _{2}(G)\geq \left\vert H^{\prime }\right\vert =\left\vert
P_{o}^{H^{\prime }}(M_{A},H^{\prime })\right\vert +\left\vert
H_{A}\right\vert +\nu (G)-\alpha _{2}(G)\geq 4(\nu (G)-\alpha _{2}(G)).
\end{equation*}
Therefore, $\frac{\nu (G)}{\alpha _{2}(G)}\leq \frac{5}{4}$.
\end{proof}

\vspace*{0.5cm}

\begin{figure}[h]
\begin{center}
\includegraphics{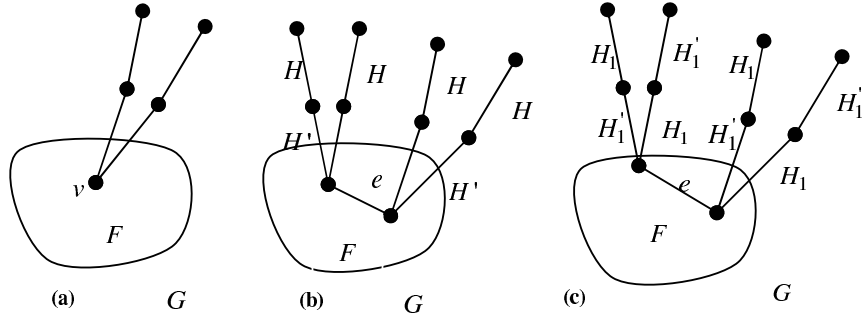}\\
\caption{$\frac{\nu(G)}{\alpha_2(G)} = \frac{5}{4}$}\label{fig1}
\end{center}
\end{figure}

\textbf{Remark 1.} We have given a proof of the theorem which is
based on the structural lemma \ref{equality}, statement \textit{(b)}
of lemma \ref{H'HpathsandPoddMAH'} and corollary \ref%
{HAbound}. It is not hard to see that the theorem can also be proved
directly using only statement \textit{(a)} of lemma
\ref{H'HpathsandPoddMAH'}. As the length of every path from $Y$ is
at least four, there are at least two edges from $H'$ lying on each
path from $Y$, therefore

\begin{center}
$\alpha _{2}(G) \geq |H'| \geq 2|Y| = 4(\nu (G)-\alpha _{2}(G))$.
\end{center}

\textbf{Remark 2.} There are infinitely many graphs $G$ for which

\begin{center}
$\frac{\nu (G)}{\alpha_2 (G)}=\frac{5}{4}$.
\end{center}

In order to construct one, just take an arbitrary graph $F$
containing a perfect matching. Attach to every vertex $v$ of $F$ two
paths of length two, as it is shown on the figure 1a.

\begin{figure}[h]
\begin{center}
\includegraphics{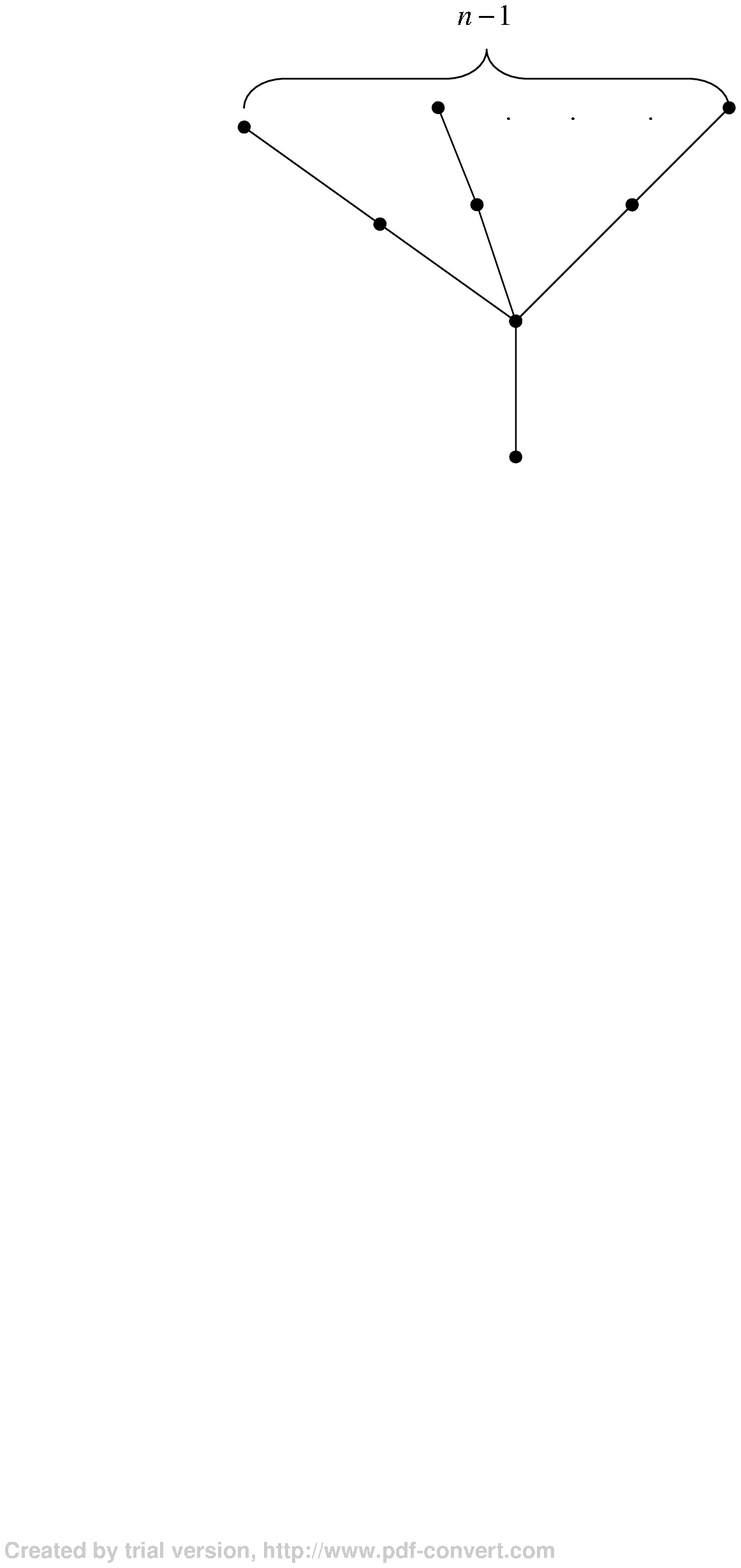}\\
\caption{$\frac{\nu(G_n)}{\lambda_2(G_n)-\alpha_2(G_n)}=n$}\label{fig2}
\end{center}
\end{figure}

Let $G$ be the resulting graph. Note that:

\begin{center}
$\nu (G)=\frac{\left\vert V(F)\right\vert }{2}+2\left\vert V(F)\right\vert =%
\frac{5\left\vert V(F)\right\vert }{2}$.
\end{center}

Let us show that for every pair of disjoint matchings $(H,H^{\prime })$
satisfying $\left\vert H\right\vert +\left\vert H^{\prime }\right\vert
=\lambda _{2}(G)$ and $e\in E(F)$ we have $e\notin H\cup H^{\prime }$. On
the opposite assumption, consider an edge $e\in E(F)$ and a pair $%
(H,H^{\prime })$ with $\left\vert H\right\vert +\left\vert H^{\prime
}\right\vert =\lambda _{2}(G)$ and $e\in H\cup H^{\prime }$. Note that
without loss of generality, we may always assume that $H$ and $H^{\prime }$
contain the edges shown on the figure 1b.

Now consider a new pair of disjoint matchings $(H_{1},H_{1}^{\prime
})$ obtained from $(H,H^{\prime })$ as it is shown on figure 1c.

Note that $\left\vert H_{1}\right\vert +\left\vert H_{1}^{\prime
}\right\vert =1+\left\vert H\right\vert +\left\vert H^{\prime }\right\vert
>\lambda_2 (G)$, which contradicts the choice of $(H,H^{\prime })$,
therefore $e\notin H\cup H^{\prime }$ and $\lambda_2 (G)=4\left\vert
V(F)\right\vert $, $\alpha_2 (G)=2\left\vert V(F)\right\vert $, hence

\begin{center}
$\frac{\nu (G)}{\alpha _{2}(G)}=\frac{5}{4}$.\\[0pt]
\end{center}

\textbf{Remark 3.} In contrast with the bound $\frac{\nu (G)}{\alpha
_{2}(G)}\leq \frac{5}{4}$, it can be shown that for every positive
integer $n\geq 2$ there is a graph $G_{n}$ such that $\frac{\nu
(G_{n})}{\lambda _{2}(G_{n})-\alpha _{2}(G_{n})}=n$. Just consider
the graph $G_{n}$ shown on the figure 2.

Note that $\nu (G_{n})=n$, $\lambda _{2}(G_{n})=n+1$, $\alpha
_{2}(G)=n$ hence

\begin{center}
$\frac{\nu (G_{n})}{\lambda _{2}(G_{n})-\alpha _{2}(G_{n})}=n$.
\end{center}

\bigskip

\bigskip

\begin{acknowledgement}
We would like to thank our reviewers for their helpful comments and
suggestions.
\end{acknowledgement}

\bigskip

\bigskip

\begin{center}
\textbf{References}
\end{center}

[1] F. Harary, Graph Theory, Addison-Wesley, Reading, MA, 1969.

[2] F. Harary, M.D. Plummer, On the core of a graph, Proc. London Math. Soc.
17 (1967), pp. 305--314.

[3] L. Lovasz, M.D. Plummer, Matching theory, Ann. Discrete Math. 29 (1986).

[4] V. V. Mkrtchyan, On trees with a maximum proper partial 0-1 coloring
containing a maximum matching, Discrete Mathematics 306, (2006), pp. 456-459.

[5] V. V. Mkrtchyan, A note on minimal matching covered graphs, Discrete
Mathematics 306, (2006), pp. 452-455.

[6] D. B. West, Introduction to Graph Theory, Prentice-Hall, Englewood
Cliffs, 1996.

\end{document}